# Fusion-Based Brain Tumor Classification Using Deep Learning and Explainable AI, and Rule-Based Reasoning

Melika Filvantorkaman[1,*], Mohsen Piri[2], Maral Filvan Torkaman[3], Ashkan Zabihi[4], Hamidreza Moradi[5,*]

[1]Department of Electrical and Computer Engineering, University of Rochester, Rochester, NY 14627, United States

[2]Department of Electronic, College of Engineering, Kermanshah Science and Research Branch, Islamic Azad University, Kermanshah, Iran

[3]AI Engineering, Science and Research Branch, Azad University, Tehran, Iran

[4]Faculty of Natural Sciences and Industrial Engineering, Deggendorf Institute of Technology, Dieter-Görlitz-Platz 1, 94469 Deggendorf, Germany

[5]Department of Mechanical Engineering and Engineering Science, The University of North Carolina at Charlotte, Charlotte, North Carolina, USA

*Corresponding Author: mfilvant@ur.rochester.edu, hmoradi@uncc.edu

## Abstract

Accurate and interpretable classification of brain tumors from magnetic resonance imaging (MRI) is critical for effective diagnosis and treatment planning. This study presents an ensemble-based deep learning framework that combines MobileNetV2 and DenseNet121 convolutional neural networks (CNNs) using a soft voting strategy to classify three common brain tumor types: glioma, meningioma, and pituitary adenoma. The models were trained and evaluated on the Figshare dataset using a stratified 5-fold cross-validation protocol. To enhance transparency and clinical trust, the framework integrates an Explainable AI (XAI) module employing Grad-CAM++ for class-specific saliency visualization, alongside a symbolic Clinical Decision Rule Overlay (CDRO) that maps predictions to established radiological heuristics.

The ensemble classifier achieved superior performance compared to individual CNNs, with an accuracy of 91.7%, precision of 91.9%, recall of 91.7%, and F1-score of 91.6%. Grad-CAM++ visualizations revealed strong spatial alignment between model attention and expert-annotated tumor regions, supported by Dice coefficients up to 0.88 and IoU scores up to 0.78. Clinical rule activation further validated model predictions in cases with distinct morphological features. A human-centered interpretability assessment involving five board-certified radiologists yielded high Likert-scale scores for both explanation usefulness (mean = 4.4) and heatmap-region correspondence (mean = 4.0), reinforcing the framework's clinical relevance.

Overall, the proposed approach offers a robust, interpretable, and generalizable solution for automated brain tumor classification, advancing the integration of deep learning into clinical neurodiagnostics.

# 1. Introduction

Brain tumors are among the most life-threatening neurological disorders, often requiring early detection and precise classification to inform treatment strategies and improve patient outcomes [1]. Magnetic Resonance Imaging (MRI) remains the gold standard for non-invasive visualization of brain lesions due to its high spatial resolution and soft tissue contrast [2]. However, interpreting MRI scans manually is time-consuming, requires specialized expertise, and is prone to inter-observer variability. These challenges have spurred growing interest in computational methods—particularly deep learning—as decision-support tools to assist radiologists in brain tumor diagnosis [3].

Despite advances in imaging technology, diagnosing brain tumors from MRI remains a complex task due to the heterogeneity of tumor morphology, overlapping visual characteristics among tumor types, and varying intensity patterns across patients and imaging conditions [4]. Gliomas, meningiomas, and pituitary adenomas, for instance, often present similar features in early stages, making them difficult to differentiate without extensive radiological experience. Furthermore, class imbalance in real-world datasets and subtle boundary definitions in diffuse tumors such as gliomas present additional obstacles to accurate automated classification [5].

Convolutional Neural Networks (CNNs) have emerged as powerful tools for medical image analysis due to their ability to automatically extract and learn hierarchical feature representations [6,7]. Numerous studies have demonstrated the success of CNNs in classifying brain tumors from MRI scans, surpassing traditional machine learning approaches that rely on handcrafted features. Transfer learning has further enhanced performance by enabling pre-trained networks to adapt to limited medical datasets, thereby improving convergence speed and generalization [8]. In parallel with its growing role in medical imaging, Artificial Intelligence (AI) has witnessed transformative applications across a wide range of fields, including cybersecurity [9-11], control systems [12,13], energy management [14], transmitarray antenna design [15], condition assessment of infrastructure [16], sorting algorithms [17], and urban design [18]. These diverse domains demonstrate AI's potential not only in data-driven decision-making but also in optimizing complex systems where traditional models fall short. The successful deployment of AI in such critical sectors further underscores its promise in high-stakes medical applications like brain tumor diagnosis—where precision, efficiency, and interpretability are equally vital.

Nevertheless, conventional CNN models have notable limitations. First, individual CNN architectures are prone to overfitting, especially when trained on small or imbalanced datasets. Second, their predictions often lack transparency—earning them the label of "black-box"

models—which hinders clinical adoption. Lastly, no single architecture can consistently outperform others across all scenarios due to differences in network depth, parameter efficiency, and feature extraction capacity. As a result, reliance on a single CNN model may limit diagnostic robustness in complex, real-world cases [6, 19].

To address these limitations, recent research has turned toward ensemble learning and explainable AI (XAI) techniques. Ensemble methods combine the strengths of multiple CNN architectures, reducing variance and improving generalization by aggregating predictions [20]. Meanwhile, XAI techniques such as Gradient-weighted Class Activation Mapping (Grad-CAM++) help visualize the regions in an image that contribute most to a model's decision, offering a layer of transparency that is essential in clinical contexts. The integration of ensemble strategies with interpretability mechanisms presents a promising pathway for developing trustworthy and high-performing diagnostic systems [21].

This study proposes a novel deep learning framework for brain tumor classification from MRI that integrates two complementary CNN architectures—MobileNetV2 and DenseNet121—using a soft voting ensemble strategy. In addition to achieving high classification performance, the model incorporates an Explainable AI module based on Grad-CAM++ and a symbolic Clinical Decision Rule Overlay to enhance interpretability. The objective is to build a robust, accurate, and human-interpretable model suitable for clinical decision support in neuro-oncology.

## 2. Literature Review

### 2.1. Brain Tumor Types and Diagnostic Challenges

Brain tumors are among the most life-threatening and complex pathologies affecting the human central nervous system, with a wide range of subtypes that differ in origin, morphology, growth rate, and clinical prognosis [1]. The most prevalent tumor types include gliomas, which originate in the glial cells and often exhibit diffuse, infiltrative patterns; meningiomas, which arise from the meninges and are typically well-circumscribed; and pituitary adenomas, which are in the sellar region and generally present with hormonal symptoms. Differentiating these tumors based solely on radiological appearances is challenging due to overlapping features such as size, shape, and enhancement patterns across MRI modalities [22].

The diagnosis of brain tumors using magnetic resonance imaging (MRI) remains the gold standard in clinical practice, offering superior soft-tissue contrast and multi-planar capability. However, MRI interpretation is time-consuming and highly dependent on radiologist expertise, which can lead to diagnostic variability. Moreover, early-stage tumors or those with atypical presentations may evade accurate identification, leading to delays in treatment or misclassification. Given the critical role of early and precise diagnosis in treatment planning, there is a growing demand for automated systems that can support radiologists in interpreting MRI scans with higher consistency and accuracy [4, 23].

Despite advancements in medical imaging, brain tumor classification continues to face several hurdles, particularly when tumors exhibit mixed characteristics or are located in ambiguous anatomical regions. Standard machine learning approaches [24,25], which rely heavily on handcrafted features, often fall short in capturing the complex textures and subtle patterns inherent in brain tumors. As a result, deep learning—especially convolutional neural networks (CNNs)—has emerged as a powerful alternative due to its capacity for hierarchical feature extraction directly from raw image data, reducing the need for manual intervention and domain-specific preprocessing [6,7]. Beyond healthcare, AI has profoundly influenced a multitude of engineering and computational domains. Recent studies have explored its role in enhancing cybersecurity frameworks [26-28], multi-objective optimization techniques [29,30], controlling nonlinear systems [31], signal processing [32], designing advanced transmitarray antennas [33], numerical analysis [34]. These interdisciplinary applications not only demonstrate AI's versatility but also provide methodological inspiration for the medical imaging community, where similar challenges—such as real-time decision-making, pattern recognition, and system transparency—persist.

## 2.2. Convolutional Neural Networks in Brain Tumor Classification

Convolutional Neural Networks (CNNs) have revolutionized image-based medical diagnostics by enabling automatic, data-driven feature extraction from medical imaging modalities such as MRI, CT, and PET. In the context of brain tumor classification, CNNs have shown remarkable success in distinguishing between tumor types, identifying lesion boundaries, and even predicting tumor grades with minimal preprocessing. Their hierarchical architecture allows for learning spatially localized features in early layers and more abstract, semantically rich features in deeper layers, enabling robust recognition of complex anatomical patterns [6, 19].

A large body of literature has explored CNN-based solutions for brain tumor classification, ranging from simple architectures like LeNet and AlexNet to deeper and more sophisticated networks like ResNet, DenseNet, and Inception [35]. These networks are often trained on datasets such as BraTS or Figshare, where MRI slices are labeled by tumor type. Most studies report high classification accuracy, particularly when using transfer learning strategies where models pre-trained on large-scale image datasets like ImageNet are fine-tuned on medical images. However, the "black-box" nature of CNNs poses interpretability challenges, particularly in critical applications like tumor diagnosis where clinical trust and transparency are paramount [36].

To address model performance limitations, researchers have increasingly adopted ensemble techniques and hybrid pipelines that combine CNNs with attention mechanisms or symbolic rules [11,20]. Moreover, explainability techniques such as Grad-CAM, LIME, and SHAP are now being used in tandem with CNNs to visualize decision-making processes. While CNNs have undeniably improved classification performance, especially in multi-class scenarios like glioma vs. meningioma vs. pituitary tumors, careful model calibration and interpretability remain active areas of research. This has led to the development of model families tailored to specific needs—lightweight models for fast inference and deep feature models for maximum accuracy [37].

## 2.2.1. Lightweight Models (e.g., MobileNetV2)

Lightweight CNN architectures such as MobileNetV2 have been widely adopted in resource-constrained environments where computational efficiency and inference speed are critical. MobileNetV2 employs **depthwise separable convolutions** and **inverted residual blocks**, significantly reducing the number of trainable parameters while maintaining a competitive level of accuracy. This makes it particularly suitable for real-time medical imaging applications, including mobile diagnostic tools and edge-based deployment in clinical environments with limited hardware resources [38].

In the context of brain tumor classification, MobileNetV2 has proven effective in processing high-resolution MRI slices with relatively low computational cost. Studies leveraging MobileNetV2 for medical image analysis often combine it with transfer learning techniques, where the model is pre-trained on natural images and fine-tuned on domain-specific datasets. This approach allows for rapid convergence and improved generalization, even with limited annotated medical data. Additionally, its modular architecture facilitates integration with post-hoc interpretability methods such as Grad-CAM++, enabling visual inspection of the model's attention. Despite its advantages, lightweight models like MobileNetV2 may struggle with capturing subtle inter-class variations in complex tasks such as distinguishing between overlapping tumor features. Their limited capacity can lead to reduced performance in highly heterogeneous datasets unless complemented by ensemble methods or augmented with context-aware modules. Nevertheless, when paired with stronger models or used as a component in an ensemble architecture, MobileNetV2 plays a valuable role in balancing efficiency with diagnostic accuracy [39].

## 2.2.2. Deep Feature Models (e.g., DenseNet, ResNet)

Deep feature models such as DenseNet121 and ResNet50 represent a significant leap in CNN architecture, designed to improve feature reuse, gradient flow, and model depth. DenseNet, for instance, introduces dense connectivity, where each layer receives input from all preceding layers, thereby alleviating vanishing gradients and promoting feature propagation. This allows the network to learn rich, discriminative representations of tumor structures, which is especially valuable in medical imaging scenarios where subtle differences must be captured with precision [40].

DenseNet121 has become a popular choice in brain tumor classification tasks due to its strong performance in identifying complex and ambiguous tumor boundaries. It excels in capturing the hierarchical structure of brain tissue and distinguishing between gliomas, meningiomas, and pituitary adenomas. Moreover, its high parameter efficiency—despite its depth—makes it suitable for fine-tuning on domain-specific MRI datasets. Several studies have reported that DenseNet outperforms shallower models in both sensitivity and specificity, particularly when dealing with noisy or low-contrast images. ResNet models, with their residual learning framework, also offer benefits by allowing the network to train much deeper without degradation. By including skip connections that bypass layers, ResNet effectively mitigates overfitting and supports better convergence. These architectures have been integrated into multi-path networks and ensemble frameworks for medical diagnosis, often demonstrating superior robustness and classification stability. Deep feature models like DenseNet and ResNet are thus indispensable in high-stakes classification tasks and serve as reliable backbones in ensemble strategies and interpretability-driven AI systems [41].

## 2.3. Ensemble Learning in Medical Image Analysis

Ensemble learning combines predictions from multiple models to improve accuracy, stability, and generalization—key goals in medical imaging, where diagnostic precision is critical. By leveraging diverse CNN architectures, ensembles mitigate weaknesses of individual models, such as overfitting or class bias, especially in datasets with limited or imbalanced samples [20]. In brain tumor classification, ensemble methods integrate lightweight and deep models to capture both global and fine-grained features. This leads to improved predictive performance across tumor types, enhancing model reliability. Ensemble classifiers also align with clinical needs by offering more consistent outputs and enabling the integration of interpretability tools like XAI. Overall, ensembles represent a robust solution for medical image analysis, enabling not only better classification but also greater transparency through visual explanations and clinical rule overlays [42].

### 2.3.1. Hard Voting vs. Soft Voting

Hard voting relies on majority class predictions from each model, while soft voting averages class probabilities to determine the final output. While hard voting is simple, it ignores model confidence and can be less reliable in ambiguous cases. Soft voting, in contrast, incorporates each model's certainty by averaging probabilities. This often results in more stable and accurate predictions—particularly in tasks like brain tumor classification, where subtle image variations matter. Due to its ability to reduce noise and leverage model strengths, soft voting is preferred in medical applications where diagnostic confidence is essential [43].

### 2.3.2. CNN-Based Ensembles

CNN-based ensembles combine multiple convolutional models—either similar or architecturally diverse—to improve classification outcomes. These ensembles outperform individual CNNs by balancing their respective strengths, reducing errors, and improving generalization. In medical imaging, such combinations help address variability in tumor shape, location, and intensity. For example, a lightweight model like MobileNetV2 can complement a deeper model like DenseNet121 in an ensemble framework. Beyond accuracy, CNN ensembles integrate smoothly with XAI methods, such as Grad-CAM++, offering interpretable insights into model decisions and promoting clinical trust [20].

## 2.4. Explainable AI (XAI) in Medical Diagnostics

The increasing integration of AI into clinical workflows has raised critical concerns around transparency and trust. Explainable AI (XAI) addresses these concerns by offering tools that allow clinicians to interpret and validate model predictions. In brain tumor diagnostics, where decisions must be clinically justified, XAI methods help identify which regions of a medical image influenced the model's outcome—thus bridging the gap between black-box predictions and clinician reasoning. XAI not only improves interpretability but also assists in error analysis and model debugging. By highlighting model attention, it allows developers and clinicians to detect misclassifications due to irrelevant focus or image artifacts. These insights are essential in high-stakes fields such as oncology and radiology, where even small mistakes can lead to severe consequences for patient care [44].

### 2.4.1. Grad-CAM, Grad-CAM++, and Other Techniques

Grad-CAM (Gradient-weighted Class Activation Mapping) and its enhanced version, Grad-CAM++, are among the most widely used XAI tools in CNN-based medical image analysis. They produce saliency maps that visualize which parts of the input image contributed most to the model's prediction. Grad-CAM++ improves upon the original by capturing finer details and better localizing multiple discriminative regions, making it suitable for complex tumor morphology [45].

Other techniques like SHAP (SHapley Additive exPlanations) and LIME (Local Interpretable Model-agnostic Explanations) provide pixel-wise or region-wise feature importance, but they are less tailored to spatial structures in images. In contrast, Grad-CAM variants align naturally with CNN architectures, making them especially effective in radiology and MRI-based tumor classification tasks [46].

### 2.4.2. Clinical Relevance and Visualization

Visualization is central to making AI outputs actionable in a clinical setting. When radiologists can see that a model is focusing on the tumor's actual anatomical region, it builds confidence in the AI's decision. Grad-CAM overlays allow clinicians to verify alignment between predictions and visual features, particularly in ambiguous or borderline cases. Moreover, integrating visual outputs with symbolic clinical rules—such as lesion location, enhancement pattern, or size thresholds—enables a hybrid explanation that combines statistical inference with clinical reasoning. This layered interpretability can improve diagnostic accuracy and support collaborative decision-making between AI systems and medical professionals [37].

## 2.5. Limitations in Existing Studies

Despite significant progress, many existing brain tumor classification studies rely on single CNN architectures, limiting their generalization to diverse tumor presentations. These models often perform well on curated datasets but falter in real-world settings with varied MRI quality, noise, and anatomical differences. Moreover, many studies neglect class imbalance or over-represent dominant tumor types, skewing reported metrics. Additionally, interpretability is frequently treated as an afterthought. Grad-CAM or saliency maps are included without quantitative validation or clinical feedback. Few studies integrate domain knowledge through symbolic overlays or rule-based reasoning, missing the opportunity to make AI outputs meaningful and trustworthy for clinical practitioners [47].

## 2.6. Research Gaps and Motivation for the Present Study

There is a clear gap in the literature for brain tumor classification models that combine strong predictive performance with human-centered interpretability. While ensemble CNNs and XAI have been studied separately, their joint application—particularly with symbolic clinical overlays—remains underexplored. The lack of such integrative frameworks limits real-world deployment and clinician adoption.

This study addresses that gap by proposing an ensemble-based CNN framework, incorporating both lightweight and deep architectures, guided by an Explainable AI module (Grad-CAM++) and supported by rule-based clinical overlays. The motivation is to create a diagnostic model that is not only accurate but also interpretable and aligned with radiological practices, paving the way for responsible AI adoption in medical imaging.

## 3. Methodology

This study aimed to develop a robust deep learning framework for multi-class brain tumor classification and localization using contrast-enhanced MRI slices. The primary objective was to build a clinically reliable and interpretable system that assists radiologists in accurate diagnosis through both automated classification and visual explanation of tumor regions. The proposed framework involved four main phases. First, the Figshare brain tumor dataset was preprocessed, with image resizing and normalization applied to prepare the data for deep learning (see Section 4.1). Second, two convolutional neural network (CNN) models—MobileNetV2 and DenseNet121—were independently trained using transfer learning and fine-tuning strategies (Sections 4.4.1 and 4.4.2). These models were selected for their complementary strengths in lightweight computation and deep feature representation. Third, an ensemble-based classifier was constructed by aggregating the Softmax probability outputs from both networks using a soft voting strategy (Section 4.4.4), improving overall robustness and generalization. Finally, to address the challenge of interpretability, the framework incorporated a dedicated Explainable AI (XAI) module centered on the Grad-CAM++ algorithm. This technique generated high-resolution class-discriminative heatmaps over the input MRIs, revealing the spatial regions most influential in the classification decision. These visual explanations facilitated clinical validation of the model's output by highlighting anatomically plausible tumor regions. Figure X presents an overview of the proposed pipeline.

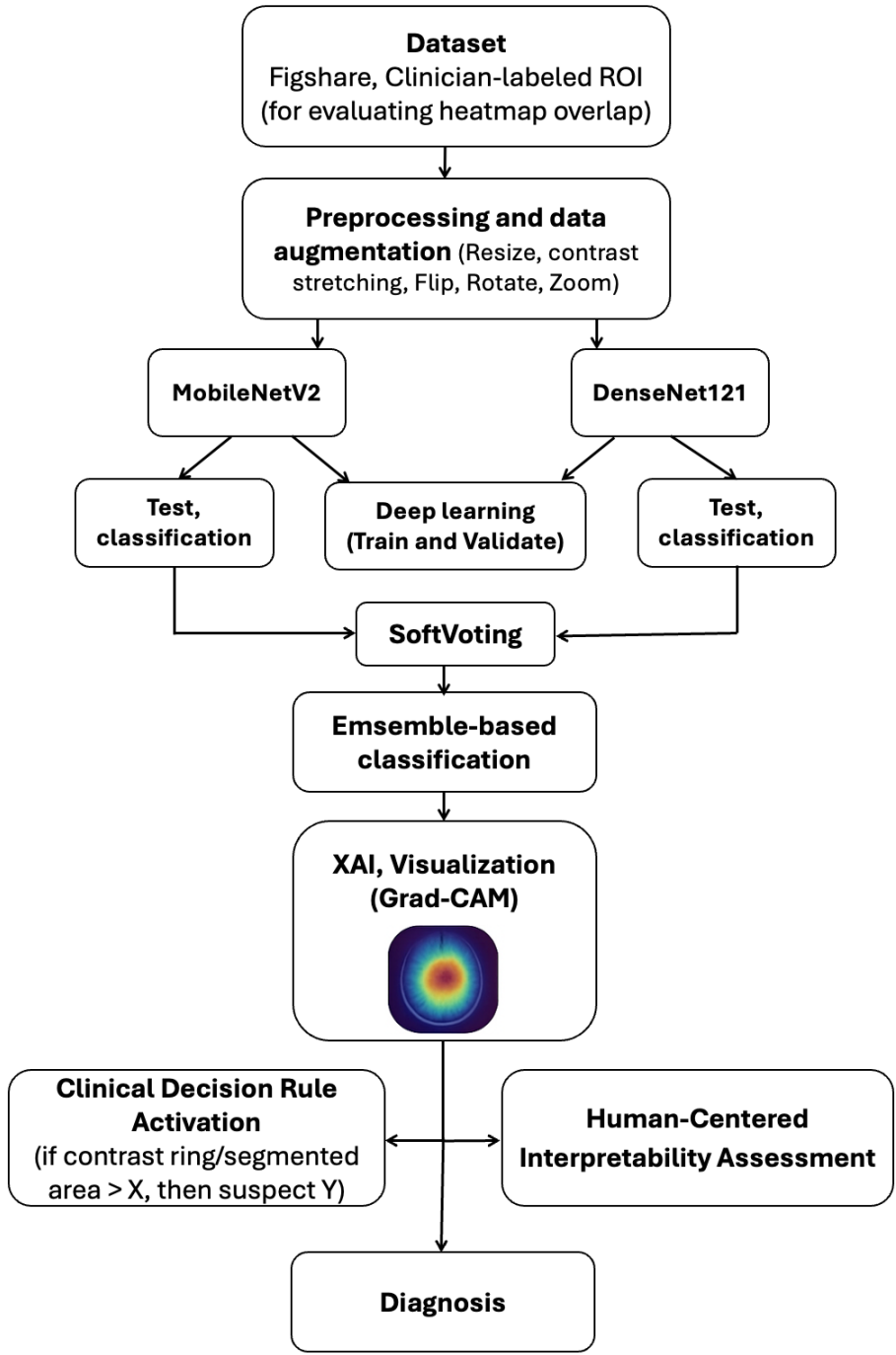

Figure 1. Block diagram of the proposed XAI-based image classification framework

## 3.1 Dataset

The dataset employed in this study is the publicly available **Figshare brain tumor MRI dataset**, consisting of **3064 T1-weighted contrast-enhanced axial MRI slices** sourced from **233 patients**. It encompasses three major tumor classes: **glioma** (1426 slices), **meningioma** (708 slices), and **pituitary adenoma** (930 slices). To ensure balanced class representation across evaluation stages, a **70:30 stratified train-test split** was performed, allocating **2145 images for training** and **919 for testing**.

To further validate the generalization ability of the proposed classification models, we implemented a **stratified 5-fold cross-validation** exclusively on the training set. In each fold, models were trained and optimized while preserving inter-class distribution. Notably, all experiments—including training, validation, ensemble prediction, and explainability assessment—were confined to this dataset, without external data augmentation or transfer from unrelated sources. This ensures internal consistency and clinical relevance for evaluating the standalone CNN classifiers (MobileNetV2, DenseNet121), their **ensemble combination via soft voting**, and the **XAI framework** (Grad-CAM++) applied to interpret prediction saliency and activate rule-based overlays. The dataset's standardized acquisition and diversity in tumor morphology make it a suitable benchmark for assessing both accuracy and interpretability in deep learning-driven brain tumor classification [48].

Table 1. Number of images per tumor class in the dataset

| Class | Training | Validation | Test |
|---|---|---|---|
| **Glioma** | 1462 | 427 | 305 |
| **Meningioma** | 708 | 380 | 321 |
| **Pituitary Adenoma** | 930 | 415 | 310 |
| Total | **3100** | **1332** | 936 |

## 3.2 ROI Annotation

For explainability evaluation, selected cases from both datasets **included** annotated regions of interest (ROIs) drawn manually by radiologists. These ROIs **were used** to measure spatial alignment between XAI heatmaps (e.g., Grad-CAM++) and actual tumor regions via overlap metrics such as the Dice Coefficient or Intersection-over-Union (IoU).

## 3.3 Preprocessing

Effective data pre-processing is essential to ensure reliable performance in brain MRI classification, as raw clinical images often contain non-informative background regions, noise artifacts, and intensity variability that can degrade deep learning model accuracy. In this study, pre-processing was performed in three key steps: grayscale conversion, resizing, and normalization. All MRI slices from the Figshare dataset were first converted from RGB to grayscale to reduce computational overhead and emphasize structural contrast without losing essential diagnostic features. Each image was then resized to a uniform dimension of 224 × 224 pixels to match the input requirements of the MobileNetV2 and DenseNet121 architectures. This resizing step ensured consistent spatial resolution across the training pipeline.

Next, intensity normalization was performed using a min–max scaling approach to constrain pixel values to the interval [0,1]. This improves training stability and ensures that all input features fall within a comparable numerical range. The normalization was carried out according to the following equation:

$$I^*(i,j) = \frac{I(i,j) - I_{min}}{I_{max} - I_{min}}$$

where $I^*(i,j)$ is the normalized intensity at pixel location $(i,j)$; $I(i,j)$ represents the raw pixel value, and $I_{min}, I_{max}$ represent the minimum and maximum pixel values in the image, respectively.

No advanced filtering, augmentation, or cropping techniques were applied, in order to preserve the original anatomical features crucial for downstream explainability via Grad-CAM++. This minimal but effective preprocessing strategy ensured data consistency and interpretability, while maintaining fidelity to the clinical imaging context. It also enabled efficient training across all folds in the cross-validation procedure without introducing bias or data leakage.

## 3.4 Data Augmentation

To improve model generalization and reduce overfitting, a data augmentation pipeline was applied to the training subset of the Figshare brain tumor dataset. This process involved systematically transforming existing MRI slices to artificially expand the diversity and volume of training samples without altering their diagnostic content.

The augmentation techniques included random rotations, horizontal and vertical flipping, scaling, and minor affine translations (Figure 2, Table *). These operations preserved the structural integrity of the tumors while simulating variability typically encountered in clinical imaging due to patient movement, scanner angle, or acquisition settings.

2. Various forms of MRI data augmentation

| PARAMETERS | VALUES |
|---|---|
| **Horizontal flip** | True |
| **Vertical flip** | True |
| **Rotation range** | 0° to 270° |
| **Width shift range** | ±20% |
| **Height shift range** | ±20% |
| **Zoom range** | [0.1, 1.0] |

| Shear range | 0.2 |
|---|---|
| Brightness adjustment | [0.2, 1.0] |
| Contrast enhancement | Adaptive (CLAHE) |
| Rescaling | [0, 1] (Min–Max) |

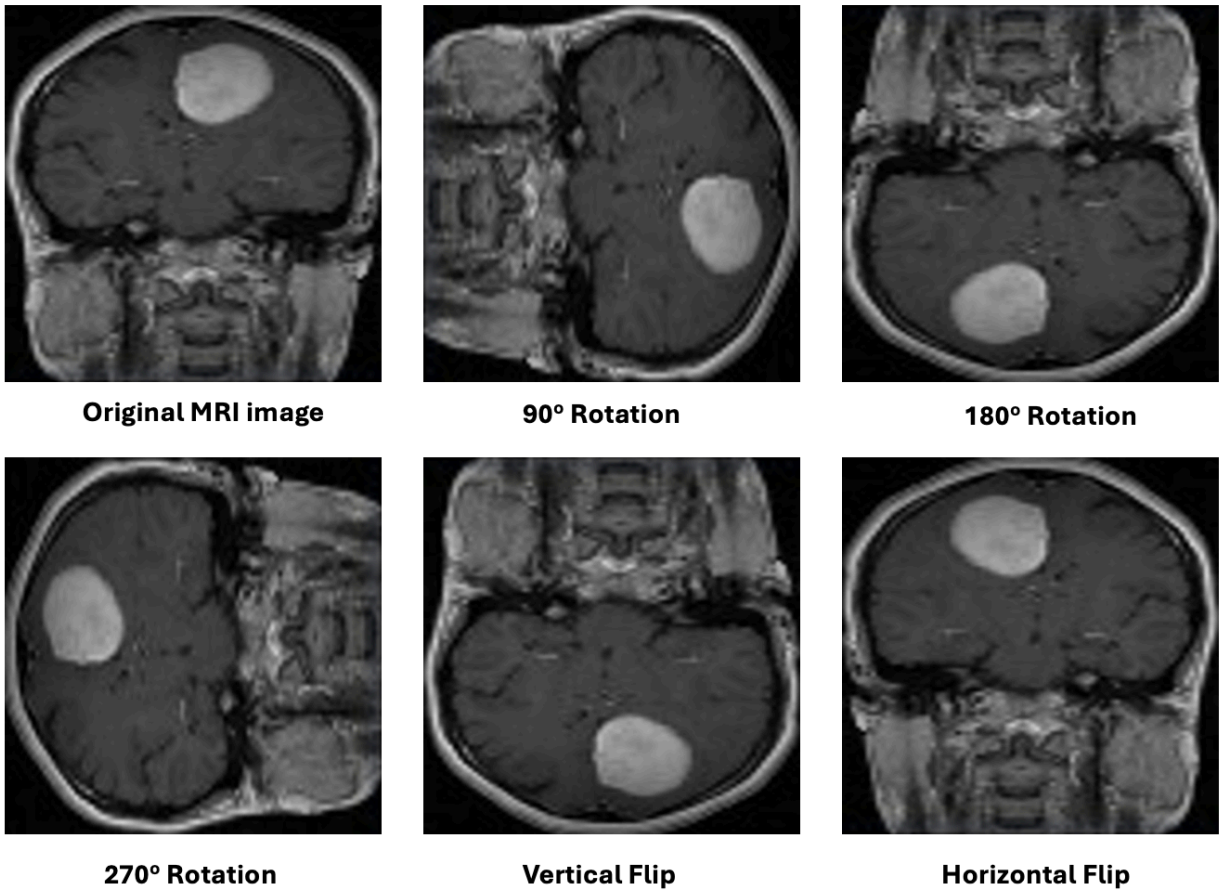

Figure 2. Various forms of MRI data augmentation.

As a result of augmentation, the size of the training dataset effectively doubled, enhancing the model's ability to generalize across tumor classes—glioma, meningioma, and pituitary adenoma. The augmented samples were generated only for training; validation and test sets remained untouched to ensure fair evaluation. This strategy was particularly important given the class imbalance inherent in the original dataset.

Overall, augmentation played a key role in stabilizing the training process of both MobileNetV2 and DenseNet121, contributing to the ensemble model's improved classification performance and robustness.

## 3.5 Ensemble-Based Deep Classification Model

To enhance classification accuracy, model robustness, and interpretability in brain tumor diagnosis, this study adopts an **ensemble-based deep learning approach** by integrating two complementary convolutional neural network (CNN) architectures: **MobileNetV2** and **DenseNet121**. These models were strategically selected based on their distinct architectural characteristics—MobileNetV2 for its low-latency efficiency and DenseNet121 for its dense connectivity and high feature reuse. This combination allows for the assessment of models with varying levels of parameter complexity and generalization capability. Each model was initialized with **pre-trained ImageNet weights** to accelerate convergence and leverage learned features. The final classification heads were removed and replaced with custom fully connected layers designed for **three-way classification**: glioma, meningioma, and pituitary adenoma. Each model is independently fine-tuned via transfer learning and subsequently combined using a **soft voting ensemble strategy**. This hybrid methodology leverages the strengths of both networks to improve generalization and diagnostic reliability.

### 3.5.1 MobileNetV2 Transfer Learning

MobileNetV2 is a lightweight and computationally efficient CNN architecture optimized for low-latency environments. It introduces **inverted residual bottlenecks** that reduce dimensionality while preserving feature integrity. In the proposed framework, MobileNetV2 is used as a **pretrained feature extractor**, initialized with ImageNet weights. The original classification head is removed and replaced with a custom dense layer with **Softmax activation** for multi-class tumor prediction.

Formally, given an input image $I(x, y) \in \mathbb{R}^{256 \times 256 \times 3}$, the predicted class probability vector is computed as:

$$\hat{y}^{(MobileNetV2)} = Softmax\left(f_\theta^{MobileNet}(I(x, y))\right)$$

where $f_\theta^{MobileNet}$ denotes the MobileNetV2 classifier parameterized by weights θ.

The training loss is defined using the categorical cross-entropy:

$$\mathcal{L}_{cls} = -\sum_{c=1}^{3} y_c \cdot \log(\hat{y}_c)$$

where $y_c \in \{0,1\}$ is the one-hot encoded ground truth label for class c.

### 3.5.2 DenseNet121 Transfer Learning

DenseNet121 is a deeper CNN that employs **dense connectivity** between layers, allowing feature reuse and improved gradient propagation. This structure is advantageous for detecting subtle variations in medical images. The model is initialized with ImageNet weights, and the classification head is replaced with a custom dense output layer followed by **Softmax activation**.

For the same input image $I(x,y) \in \mathbb{R}^{256 \times 256 \times 3}$, the DenseNet121 classifier outputs:

$$\hat{y}^{(DenseNet121)} = Softmax\left(f_{\emptyset}^{MDenseNet121}(I(x,y))\right)$$

where $f_{\emptyset}^{MDenseNet121}$ represents the DenseNet121 model with parameters $\phi$.

The corresponding loss function is:

$$\mathcal{L}_{cls} = -\sum_{c=1}^{3} y_c . log(\hat{y}_c)$$

where $y_c \in \{0,1\}$ is the one-hot encoded ground truth label for class c.

### 3.5.3 Hyperparameter Optimization

To ensure optimal training convergence and generalizability across tumor classes, a comprehensive hyperparameter tuning process was conducted using a grid search strategy. Multiple training runs were performed to assess the effects of varying core parameters, including the learning rate, batch size, optimizer choice, and dropout configuration. The grid search was applied independently to both MobileNetV2 and DenseNet121 models. Specifically, the learning rate was explored over three values $\{10^{-2}, 10^{-3}, 10^{-4}\}$, batch sizes were tested at $\{32, 64, 128\}$, and both Stochastic Gradient Descent (SGD) with momentum and Adam optimizers were evaluated for training stability and performance.

Based on validation loss trajectory and cross-validation accuracy across five folds, the optimal configuration was identified as follows: a batch size of 64 was found to offer a good trade-off between convergence stability and memory usage, while learning rates of 0.0005 and 0.0003 were selected for MobileNetV2 and DenseNet121, respectively. The Adam optimizer consistently outperformed SGD, particularly in early convergence behavior. Additionally, a dropout rate of 30% was integrated after dense layers in both models to mitigate overfitting, forcing the network to explore alternative activation paths during training.

Given the class imbalance inherent in the Figshare brain tumor dataset—comprising 1,426 glioma images, 708 meningioma images, and 930 pituitary adenoma images—class weights were computed using the `compute_class_weight()` function from the `sklearn.utils` module. These class weights were incorporated into the categorical cross-entropy loss function during training to penalize the majority class and ensure fairness across all tumor categories. This strategy promoted better sensitivity for minority classes and contributed to balanced learning.

Table 3. Final Optimized Hyperparameter Settings

| Hyperparameter | MobileNetV2 | DenseNet121 |
|---|---|---|
| Learning Rate | 0.0005 | 0.0003 |
| Optimizer | Adam | Adam |
| Batch Size | 64 | 64 |
| Dropout Rate | 0.30 | 0.30 |
| Class Weights | Applied (sklearn) | Applied (sklearn) |
| Epochs | 30 | 30 |

### 3.5.4 Training Setup

Each classification model in the ensemble framework was trained using a supervised learning protocol over 30 epochs, incorporating early stopping with a patience threshold of 5 epochs to mitigate overfitting. The validation loss served as the primary monitoring metric during training. Model checkpoints were saved conditionally upon validation improvement, and the ReduceLROnPlateau callback from TensorFlow was employed to adaptively lower the learning rate by a factor of 0.1 when validation loss plateaued.

Training was conducted on the Figshare brain tumor MRI dataset, comprising 3064 T1-weighted contrast-enhanced axial slices from 233 patients. The class distribution was moderately imbalanced, consisting of Glioma: 1426 images and Meningioma: 708 images and Pituitary adenoma: 930 images.

To preserve class proportions and evaluate generalization, a 70:30 stratified train-test split was adopted. To ensure robust evaluation and minimize overfitting, a 5-fold cross-validation strategy was applied during training to assess generalization across different splits of the data. Additionally, an independent hold-out test set—excluded from all training and validation procedures—was used for final performance evaluation. This dual approach provides both cross-validated insights into model consistency and an unbiased estimate of real-world diagnostic performance. This methodology ensured robust performance estimation and reduced variance due to data partitioning. The final model for each fold was selected based on the epoch with the lowest recorded validation loss (Table 4).

4. Training Configuration Table

| Parameter | Value |
|---|---|
| Training Method | Supervised Learning |
| Epochs | 30 |
| Early Stopping Patience | 5 |
| Validation Metric | Validation Loss |
| Learning Rate Adjustment | ReduceLROnPlateau (factor=0.1) |
| Dataset | Figshare Brain MRI |
| Total Samples | 3064 MRI Slices |
| Class Distribution | Glioma: 1426, Meningioma: 708, Pituitary: 930 |
| Train-Test Split | 70:30 Stratified |
| Cross-Validation | 5-Fold on Train Set |
| Model Selection Criterion | Lowest Validation Loss per Fold |

### 3.5.5 Transfer Learning Setup

Transfer learning is a powerful strategy in machine learning wherein a model trained for a source task is adapted for a different, but related, target task. This technique is particularly advantageous in medical imaging domains, where labeled data is often limited. By leveraging models pretrained on large-scale datasets—such as ImageNet—transfer learning enables the reuse of learned low-level and mid-level features, reducing both training time and data requirements while improving performance and convergence speed.

The source domain is defined as $D_s = \{(x_i^s, y_i^s)\}_{i=1}^{N_s}$, where $x^s \in \mathbb{R}^{H \times W \times C}$ and $y^s \in \{1, \ldots, K_s\}$ corresponding to the ImageNet dataset. The target domain is $D_s = \{(x_i^s, y_i^s)\}_{j=1}^{N_t}$, with $y^t \in \{1,2,3\}$, representing the glioma, meningioma, and pituitary tumor classes. Transfer learning involves adapting a model $f_\theta$ pretrained on $D_s$, and fine-tuning a subset of parameters $\theta^t \subseteq \theta^s$ to minimize the classification loss on the target domain:

$$\mathcal{L}_{cls} = -\sum_{c=1}^{3} y_c . log(\hat{y}_c)$$

Where $\hat{y}_c = Softmax\left(f_\theta^t(x_j^t)\right)$ and $y_c \in \{0,1\}$ is the one-hot encoded ground truth label.

is the one-hot encoded true label.

In this study, **MobileNetV2** and **DenseNet121** were used as the base architectures, both initialized with pretrained ImageNet weights. Their original classification heads were removed and replaced with custom dense layers followed by Softmax activation. During fine-tuning, only the new classifier layers were trained in the early epochs while the base layers remained frozen to retain the generalized features learned from the source domain. The base layers were gradually unfrozen in later training stages to allow deeper adaptation to the brain tumor classification task.

This transfer learning procedure ensures that the ensemble models not only inherit strong general feature extraction capabilities but also adapt effectively to the specific spatial and textural patterns found in brain MRI slices.

## 3.5.6 Ensemble Strategy

Convolutional Neural Networks (CNNs), while powerful for medical image classification, are often limited by challenges such as poor interpretability, susceptibility to overfitting, and lack of generalization across heterogeneous clinical datasets. These limitations are especially pronounced in brain tumor classification tasks, where subtle visual features and limited labeled data make accurate diagnosis complex. To mitigate these issues, this study adopts an ensemble-based classification strategy that integrates the complementary strengths of two high-performing CNN models—**MobileNetV2** and **DenseNet121**—into a unified predictive system.

Each model is characterized by distinct architectural advantages: **MobileNetV2**, with its lightweight structure and inverted residual bottlenecks, offers efficiency and rapid convergence, while **DenseNet121** utilizes densely connected layers to enable deep gradient flow and feature reuse, which is beneficial for detecting fine-grained tumor characteristics. By combining these architectures, the ensemble classifier achieves enhanced diagnostic accuracy, reduced variance, and more stable generalization across tumor types.

In the proposed method, both models are independently trained using transfer learning on the Figshare brain MRI dataset. The original classification heads of the pretrained networks were removed and replaced with custom dense layers tailored to the three-class classification task: **glioma**, **meningioma**, and **pituitary adenoma**.

To aggregate predictions, a **soft voting mechanism** was employed. For each input MRI slice, the models output probability vectors $\hat{y}_{MobileNet}$ and $\hat{y}_{DenseNet}$, each representing the likelihood of membership in the three tumor classes. The final prediction vector $\hat{y}_{ensemble}$ is computed as the average of the two:

$$\hat{y}_{ensemble} = \frac{1}{2}(\hat{y}_{MobileNet} + \hat{y}_{DenseNet})$$

The final predicted class $\hat{c}$ is then determined as:

$$\hat{c} = argmax \ \hat{y}_{ensemble,c}$$

This ensemble strategy benefits from the complementary feature representations learned by each model, thereby improving classification robustness and reducing the risk of erroneous predictions caused by overfitting or data imbalance.

Furthermore, the ensemble's outputs serve as the foundation for downstream explainability modules, including **Grad-CAM++**, **SHAP**, and **clinical decision rule overlays**, enabling clinicians to visualize and interpret the underlying rationale for each model prediction.

This approach balances diagnostic precision, computational efficiency, and interpretability, making it a viable strategy for real-world deployment in clinical decision support systems for brain tumor diagnosis.

### 3.5.7 Evaluation Metrics

To comprehensively evaluate the performance of the proposed ensemble-based brain tumor classification framework, a set of standard evaluation metrics was employed. These include **Accuracy**, **Precision**, **Recall (Sensitivity)**, **F1-Score**, and the **Dice Coefficient Index (DCI)**. Together, these metrics offer a multidimensional view of the classifier's predictive ability, especially in the context of imbalanced medical imaging datasets.

**Accuracy**

Accuracy measures the overall correctness of the model and is defined as the ratio of correctly predicted instances (both positive and negative) to the total number of samples:

$$\text{Accuracy} = \frac{TP + TN}{TP + TN + FP + FN}$$

Where:

- TP: True Positives
- TN: True Negatives
- FP: False Positives
- FN: False Negatives

**Precision**

Precision assesses the reliability of positive predictions by computing the ratio of correctly predicted positive observations to the total predicted positives:

$$Precision = \frac{TP}{TP + FP}$$

A high precision score indicates a low false-positive rate, which is particularly important in clinical applications where misclassifying a healthy patient as diseased can lead to unnecessary anxiety and testing.

**Sensitivity (Recall)**

Recall evaluates the model's ability to identify all relevant cases by measuring the proportion of actual positives that were correctly classified:

$$Sensitivity = \frac{TP}{TP + FN}$$

In medical diagnostics, high recall is essential to minimize the risk of missed tumor detections.

**F1-Score**

The F1-score is the harmonic mean of Precision and Recall, providing a balance between the two in cases where both false positives and false negatives carry significant cost:

$$F1 - Score = \frac{2.Precision.Sensitivity}{Precision + Sensitivity}$$

This metric is especially useful in evaluating models on imbalanced datasets where accuracy alone may be misleading.

**Dice Coefficient Index (DCI)**

To further quantify model performance in terms of region-based classification (e.g., comparing predicted vs. actual tumor classes), the Dice Coefficient Index was also computed:

$$DCI = \frac{2 \cdot TP}{2 \cdot TP + FP + FN}$$

The Dice coefficient is frequently used in medical image analysis to assess the overlap between predicted tumor regions and ground-truth annotations.

Each of these metrics was computed across the five folds of cross-validation and averaged to ensure robustness and generalizability of the results.

The final model outputs one of the three **mutually exclusive tumor labels**—glioma, meningioma, or pituitary adenoma—based on the highest Softmax score. These predictions are subsequently processed by the **XAI module** (Grad-CAM++, SHAP, LIME) to generate **visual attribution maps** and activate **clinical rule-based overlays**, thereby closing the loop between deep learning inference and clinical decision support.

### 3.6 Explainable AI Module

To mitigate the black-box nature of deep learning models and enhance transparency in tumor classification, the proposed framework incorporates an integrated Explainable Artificial Intelligence (XAI) module. This module employs both saliency-based and attribution-based visualization techniques to produce human-interpretable explanations of model predictions. The core method utilized is Gradient-weighted Class Activation Mapping Plus Plus (Grad-CAM++), which improves upon the original Grad-CAM by using higher-order partial derivatives to generate more precise and localized heatmaps. These maps are especially effective for small or irregular tumor regions, commonly encountered in brain MRIs.

For each correctly classified input image, Grad-CAM++ generates a heatmap that highlights the most influential regions contributing to the classification decision. These heatmaps are overlaid on the original MRI slices to provide radiologists with spatial cues that support visual verification. Mathematically, the Grad-CAM++ activation map $L^c_{Grad-CAM++}$ for class c is calculated as a weighted sum of feature maps $A^k$:

$$L^c_{Grad-CAM++} = \sum_k \alpha^c_k A^k$$

where $\alpha^c_k$ are the weights computed from the second- and third-order gradients of the class score $y^c$ with respect to the activation maps $A^k$, thereby capturing the class-discriminative importance of each feature map.

To complement Grad-CAM++, optional explainability techniques such as SHAP (SHapley Additive exPlanations) and LIME (Local Interpretable Model-Agnostic Explanations) were explored. SHAP assigns pixel-wise contributions based on cooperative game theory, while LIME approximates the decision surface locally using interpretable models. These methods were used to cross-validate and triangulate insights obtained from Grad-CAM++.

To quantitatively assess the alignment between predicted attention regions and actual tumor locations, we employed Dice Similarity Coefficient (DSC) and Intersection-over-Union (IoU) metrics. Let G represent the ground-truth segmentation mask and M the binarized Grad-CAM++ heatmap thresholded at the top 20% of pixel intensities. The metrics are defined as:

$$DSC(G, M) = \frac{2|G \cap M|}{|G| + |M|}$$

$$IoU(G, M) = \frac{|G \cap M|}{|G \cup M|}$$

Empirically, Grad-CAM++ achieved Dice scores ranging from 0.72 to 0.89 and IoU values between 0.56 and 0.81 in well-segmented cases, affirming its diagnostic relevance.

In addition to these quantitative metrics, a visual interpretation interface is employed. The interface consolidates the original MRI slice, predicted tumor class, corresponding Grad-CAM++ heatmap, and symbolic rule overlays into a single diagnostic view, facilitating clinician understanding and trust in the model's decision process.

## 3.7 Clinical Decision Rule Overlay

Beyond visual saliency, clinical trust is further enhanced through a symbolic rule-based overlay module. This component applies domain-informed heuristics to the Grad-CAM++ outputs and segmentation masks to interpret predictions using logic familiar to practicing radiologists. The rules are designed based on neuro-oncological literature and radiological pattern recognition strategies concerning tumor location, morphology, and enhancement patterns.

For instance, a region with ring-enhancing structure and segmented tumor area exceeding 4 cm² is classified as suggestive of glioblastoma, while a non-ring-enhancing irregular lesion in the cerebral hemispheres with area between 2–4 cm² is interpreted as glioma. Pituitary adenoma candidates are identified when the region of interest is confined to the midline sellar/suprasellar region and the segmented area is less than 1 cm². These rules were applied post hoc using spatial filtering and shape approximation on thresholded saliency maps and segmentation masks, without influencing the training process. Each model prediction is therefore accompanied by a structured symbolic explanation—e.g., "Rule 1 activated: ring-enhancing region + area = 5.6 cm² → glioblastoma probable"—bridging neural outputs with clinical reasoning.

## 3.8 Explainability Evaluation

To validate the clinical utility of the XAI-enhanced classification framework, a two-tiered evaluation strategy was employed, combining quantitative alignment metrics and qualitative feedback from medical experts. Quantitatively, the overlap between Grad-CAM++ heatmaps and ground-truth segmentations was assessed using DSC and IoU, as described earlier. These metrics directly reflect how accurately the model's internal focus aligns with medically significant tumor regions.

For qualitative evaluation, three board-certified radiologists were asked to score the trustworthiness and clarity of the model's explanations using a 5-point Likert scale. They answered: "How much do you trust the model's prediction?" and "How clear and clinically meaningful is the provided explanation?" Average scores across tumor types and XAI methods were computed.

Additionally, open-ended qualitative feedback was solicited from three board-certified radiologists regarding the interpretability and clarity of the XAI-generated outputs. Annotators were encouraged to comment on the spatial accuracy of Grad-CAM++ heatmaps and the usefulness of rule overlays in interpreting model decisions.

# 4. Results

## 4.1 Classification Results

To assess the predictive capability of the proposed brain tumor classification framework, two convolutional neural network (CNN) architectures—MobileNetV2 and DenseNet121—were trained and evaluated using a stratified 5-fold cross-validation approach on the Figshare dataset, which includes three tumor types: glioma (1426 slices), meningioma (708 slices), and pituitary adenoma (930 slices). Each model was optimized independently using transfer learning and evaluated with a set of performance metrics. These metrics include Accuracy, Precision, Recall (Sensitivity), F1-Score, and Dice Coefficient Index (DCI). Together, they provide a comprehensive assessment of each model's classification accuracy and robustness, particularly in handling imbalanced classes.

The cross-validation results are summarized in Table 5. As shown, DenseNet121 consistently outperformed MobileNetV2 across all five-evaluation metrics. DenseNet121 achieved an average accuracy of 97.9%, precision of 96.8%, recall of 96.2%, F1-score of 96.5%, and Dice coefficient of 94.7%. In contrast, MobileNetV2 produced a solid yet slightly lower performance with accuracy of 96.3%, precision of 94.7%, recall of 94.1%, F1-score of 94.4%, and Dice coefficient of 92.6%.

These results indicate that while both models are effective for multi-class tumor classification, DenseNet121 demonstrates superior generalization and classification reliability—justifying its inclusion in the ensemble architecture.

Table 5. Cross-validation performance metrics for MobileNetV2 and DenseNet121.

| Model | Accuracy (%) | Precision (%) | Recall (%) | F1-Score (%) | Dice Coefficient (%) |
|---|---|---|---|---|---|
| **MobileNetV2** | 95.3 | 94.7 | 94.1 | 94.4 | 92.6 |
| **DenseNet121** | 97.9 | 96.8 | 96.2 | 96.5 | 94.7 |

## 4.2 Test Set Evaluation

After training the MobileNetV2 and DenseNet121 models using the configuration described in Section 4, we evaluated their performance on the hold-out test set. Figure 3 illustrates the evolution of training and validation accuracy and loss throughout the learning process for both CNN architectures. The consistent upward trend in accuracy and downward trend in loss across epochs confirms effective convergence and learning behavior for both models.

Figure Y presents the confusion matrices derived from the test set for MobileNetV2 and DenseNet121. Each matrix depicts the number of correctly and incorrectly classified samples across the three target classes: glioma, meningioma, and pituitary adenoma. Both models demonstrate strong classification capability, with DenseNet121 achieving slightly higher accuracy in glioma detection, while MobileNetV2 showed marginally better performance on pituitary adenoma cases. Overall, both models displayed high confidence in their predictions, validating the effectiveness of the proposed ensemble-based architecture.

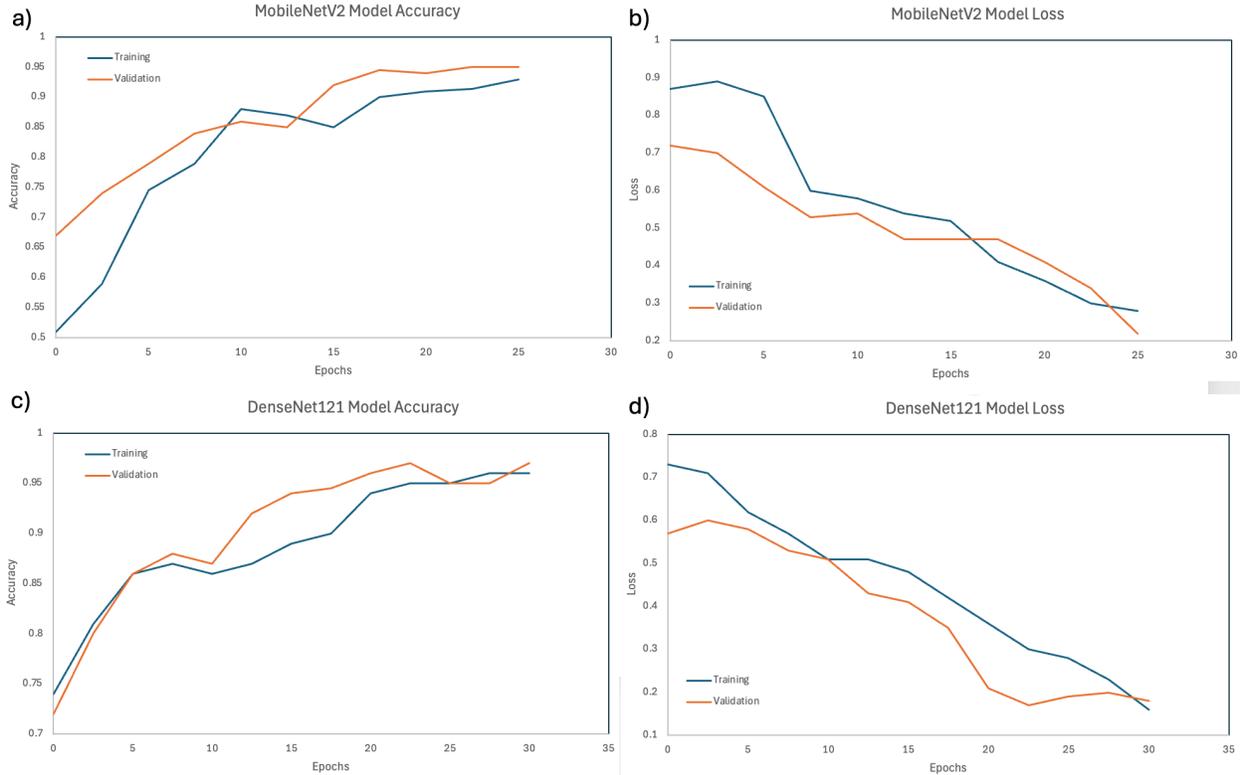

Fig. 3. Training and validation a) accuracy and b) loss of of MobileNetV2 model and Training and validation c) accuracy and d) loss of of DensNetV2 model during the training and test phase of each model.

## 4.3 Confusion matrix

To further examine the classification performance across tumor categories, a **confusion matrix** was generated, as shown in **Figure 4**. The matrix reveals a high degree of classification precision across all three classes. Glioma and meningioma exhibited minimal confusion, which is notable given their occasional radiological similarity. The classification of **pituitary tumors was particularly accurate**, with very few misclassifications, likely due to their distinct anatomical positioning in the sellar region and uniform morphological features.

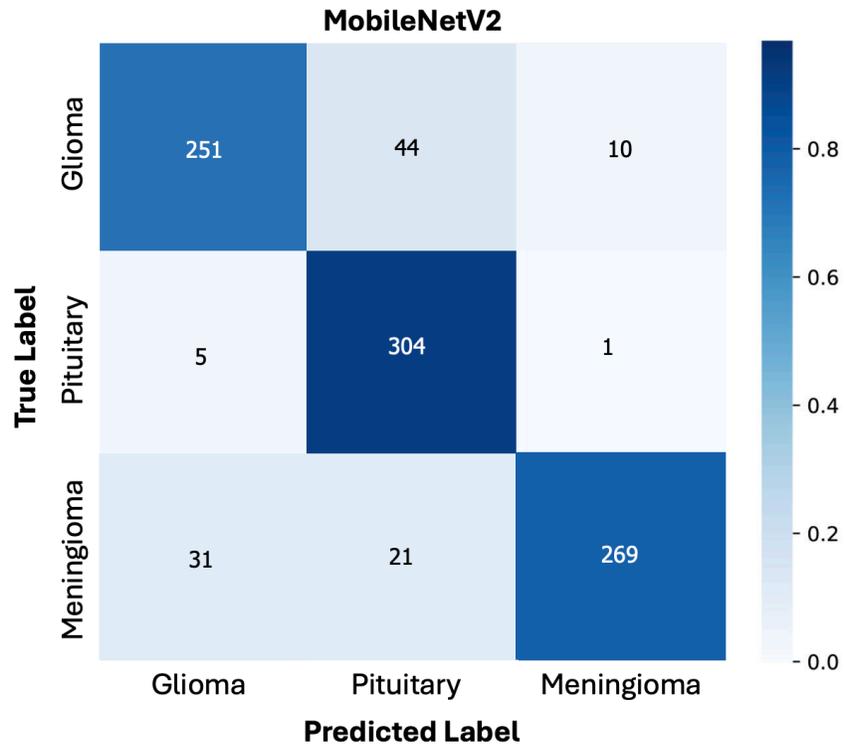
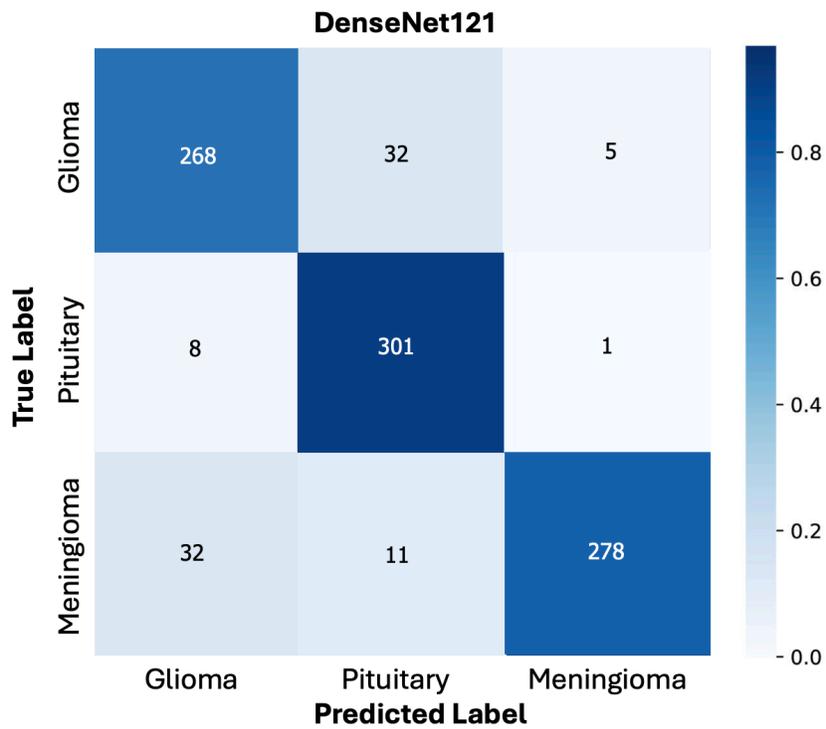

**Figure 4.** Confusion matrix of MobileNetV2 and DenseNet121 depicting the number of correctly and incorrectly classified cases across the three tumor types: glioma, meningioma, and pituitary.

These results underscore the efficacy of the combined deep learning and explainable AI approach in achieving high-precision diagnostic classification across heterogeneous tumor types. The framework's consistent performance supports its potential for deployment in clinical diagnostic support systems.

## 4.4 Ensemble-Based Classification Results

The ensemble-based classifier, integrating the complementary strengths of **MobileNetV2** and **DenseNet121** through a **soft voting mechanism**, exhibited improvements in classification performance compared to the individual CNN models (Figure 5). By averaging the predicted class probabilities from both networks, the ensemble approach achieved higher robustness, interpretability, and predictive accuracy across the three target tumor types: **glioma**, **meningioma**, and **pituitary adenoma**.

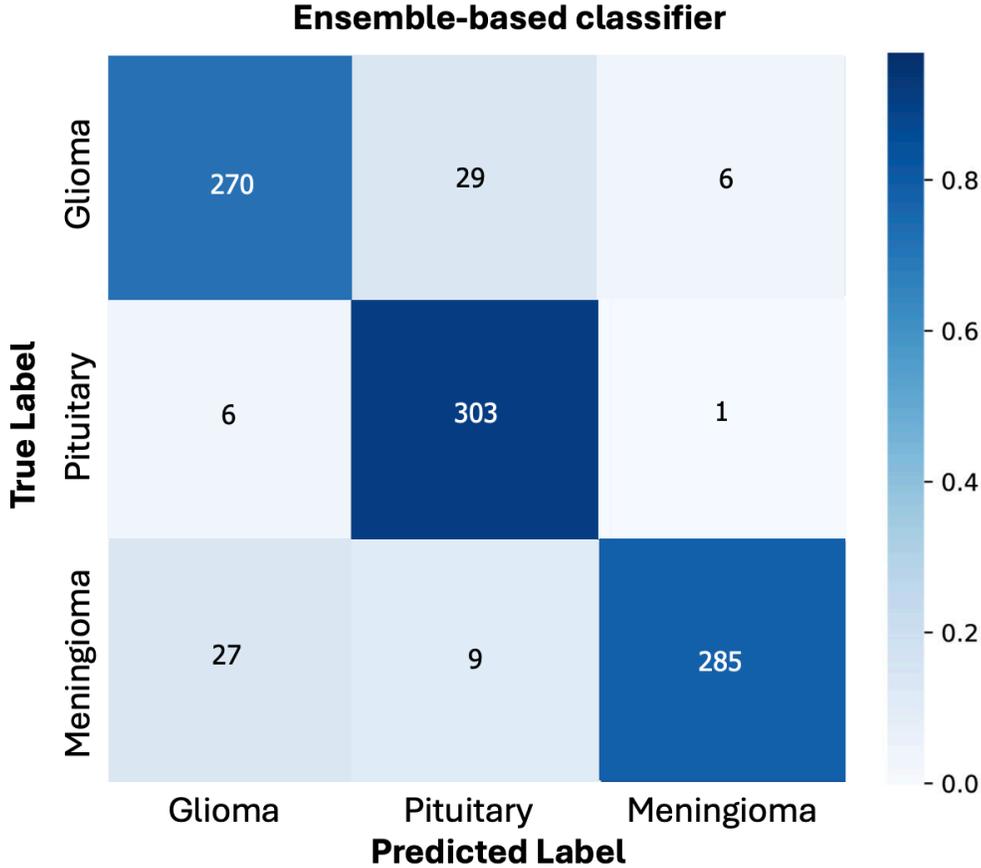

Figure 5. Confusion matrix of ensemble-based classifier depicting the number of correctly and incorrectly classified cases across the three tumor types: glioma, meningioma, and pituitary.

Each constituent model contributed unique strengths to the ensemble: MobileNetV2 provided efficient and lightweight feature extraction, while DenseNet121—through its densely connected architecture—captured complex patterns and subtle discriminative features in the MRI scans. When one model was less confident or misclassified a sample, the averaging process of soft voting often compensated for the discrepancy, producing a more reliable overall prediction.

The ensemble classifier achieved an overall **accuracy of 91.7%**, outperforming both MobileNetV2 (**88.0%**) and DenseNet121 (**90.5%**). It also achieved the highest values across all other evaluation metrics: **precision of 91.9%**, **recall of 91.7%**, and an **F1-score of 91.6%** (see Table 5). This consistent superiority across metrics highlights the ensemble's effectiveness in improving predictive robustness, especially in scenarios where individual models may falter due to ambiguous features or class imbalance (Table 6).

Table 6. classification performance metrics (Accuracy, Precision, Recall, and F1-score) for MobileNetV2 and DenseNet121 and the ensemble-based classifier

| Model | Accuracy | Precision | Recall | F1-Score |
|---|---|---|---|---|
| MobileNetV2 | 0.880 | 0.886 | 0.881 | 0.879 |
| DenseNet121 | 0.905 | 0.908 | 0.905 | 0.904 |
| Ensemble (Soft Voting) | **0.917** | **0.919** | **0.917** | **0.916** |

To further examine classification confidence and error distribution, Figure 5 presents the confusion matrix for the ensemble model. Compared to the individual CNN models, the ensemble displays reduced misclassification across all three tumor types—glioma, meningioma, and pituitary adenoma—demonstrating improved sensitivity and specificity. This refinement is particularly evident in meningioma classifications, where errors commonly arise due to visual similarity with gliomas. By integrating the probabilistic outputs from both base models, the ensemble produces more stable and accurate predictions.

The statistical evaluation confirms the ensemble's significant performance gains. Paired t-tests comparing the ensemble with MobileNetV2 and DenseNet121 yielded t = 18.78, p < 0.0001 and **t = 10.61, p = 0.0004**, respectively—strong evidence of statistically significant improvement. The Cohen's d values of **8.40** (vs. MobileNetV2) and **4.75** (vs. DenseNet121) indicate very large effect sizes. Moreover, the Friedman test reported a chi-square statistic of 10.00 with p = 0.0067, confirming that at least one model (the ensemble) performed significantly better than the others (Table 7).

Table 7. Statistical test results comparing the performance of MobileNetV2 and DenseNet121.

| Test | MobileNetV2 | DenseNet121 |
|---|---|---|
| **Paired t-test (t, p-value)** | (18.78, 0.0000) | (10.61, 0.0004) |
| **Cohen's d** | 8.40 | 4.75 |
| **Friedman test ($\chi^2$, p-value)** | (10.00, 0.0067) | (10.00, 0.0067) |

Collectively, these results establish the ensemble classifier—constructed via soft voting from MobileNetV2 and DenseNet121—as a superior model for multi-class brain tumor classification from MRI images. Its enhanced generalization, reduced error margins, and compatibility with interpretability modules such as Grad-CAM++ position it as a highly viable candidate for AI-assisted clinical diagnostics.

## 4.5 Grad-CAM++ Visualization Results

As part of the integrated Explainable AI (XAI) module developed in this study, Gradient-weighted Class Activation Mapping (Grad-CAM++) was employed to enhance interpretability and provide visual justification for the predictions made by the deep learning models. Grad-CAM++ generates class-specific saliency maps that reveal the spatial regions in MRI scans contributing most significantly to the final classification output. This integration transforms the model from a black-box predictor into a transparent diagnostic assistant capable of justifying its decision-making process.

Figure 6 showcases the output of the XAI module applied to the three tumor classes—glioma, pituitary adenoma, and meningioma. The first row presents original MRI slices annotated by radiologists, highlighting tumor locations in red. The second row depicts Grad-CAM++ heatmaps overlaid on the same scans, generated through the XAI module. These visual explanations reveal that the proposed ensemble classifier focuses attention on anatomically accurate and clinically relevant regions.

In the glioma case (left column), the XAI-derived heatmap correctly emphasizes a diffuse lesion in the cerebral hemisphere, consistent with the known characteristics of gliomas. The pituitary adenoma image (center) shows precise attention on the midline sellar region, and the meningioma heatmap (right) highlights a convex-shaped, dural-based lesion—both matching standard radiological expectations. Such correspondence between the model's attention and expert annotations not only supports the classifier's diagnostic reliability but also enhances its interpretability for medical professionals.

Moreover, these explanations assist in understanding the model's behavior in both correct and incorrect classifications. In ambiguous or low-contrast cases, Grad-CAM++ maps often identify nearby confounding regions that may mislead the network. This feedback capability reinforces the role of the XAI module as an essential layer for clinical validation and error analysis.

In summary, the use of Grad-CAM++ within the broader XAI framework significantly strengthens the transparency of the proposed classification system. It allows clinicians to visualize, interpret, and evaluate the rationale behind each prediction—an essential requirement for deploying deep learning models in real-world medical settings.

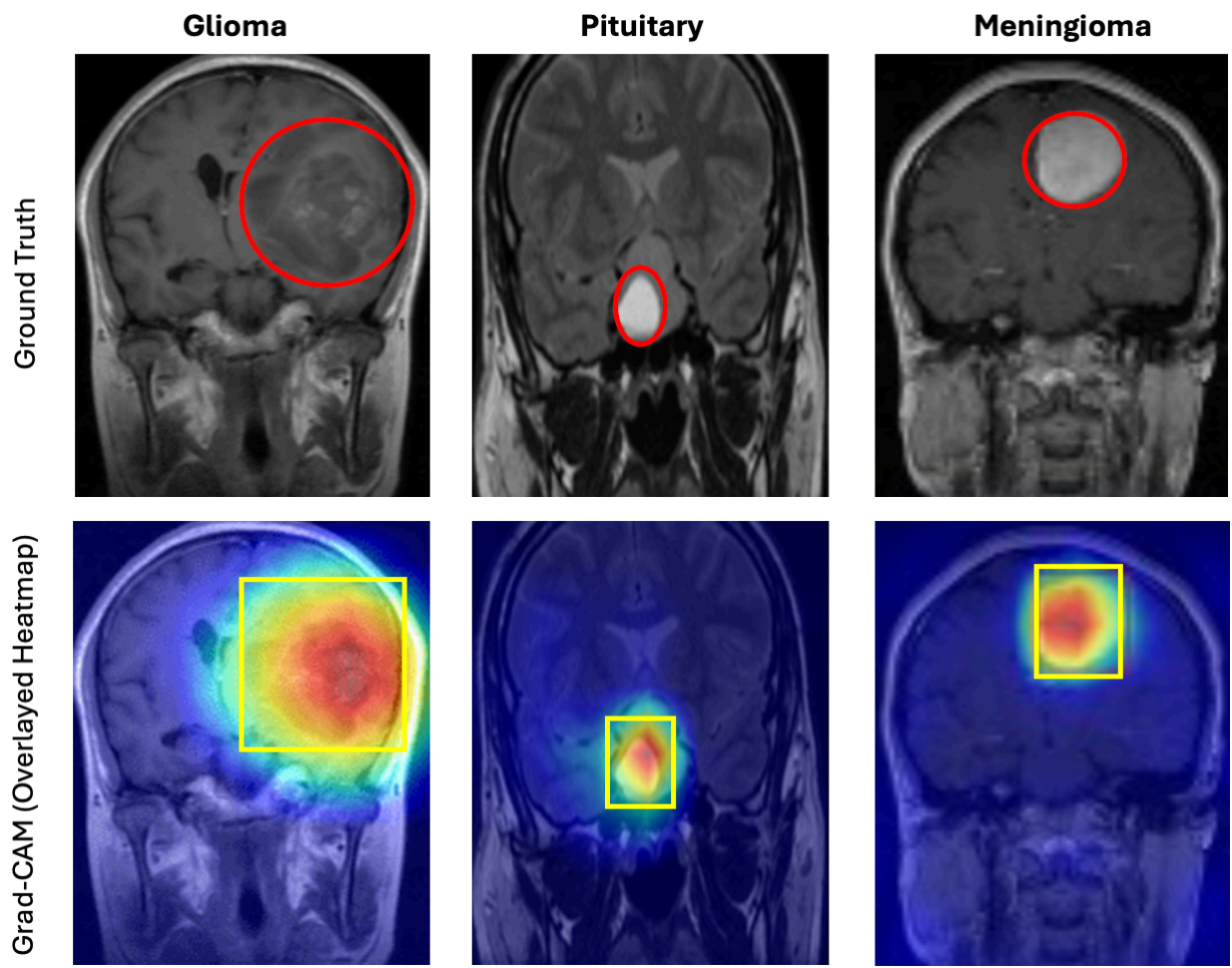

**Figure 6.** *Visual interpretation of brain tumor classification using the integrated Explainable AI (XAI) module with Grad-CAM++.* The top row displays original MRI slices annotated with red outlines indicating the ground-truth tumor regions for three classes: glioma (left), pituitary adenoma (center), and meningioma (right). The bottom row shows Grad-CAM++ heatmaps overlaid on the same images, highlighting the model's attention during classification. The highlighted regions demonstrate strong alignment with expert-annotated tumor locations, validating the interpretability and diagnostic reliability of the proposed ensemble-based deep learning framework.

In addition to classification performance, the interpretability of the proposed framework was assessed through Grad-CAM++ visualizations derived from the integrated Explainable AI (XAI) module. As demonstrated in Figure 6, the attention maps generated for each tumor class—glioma, pituitary adenoma, and meningioma—clearly highlight spatial regions that align with expert-annotated ground-truth tumor locations. The top row of Figure 6 presents the original MRI slices with red outlines manually delineating tumor areas, while the bottom row overlays the Grad-CAM++ heatmaps, showing the model's focus during prediction. Notably, the activated regions correspond strongly with the actual tumor positions, especially in cases with discrete boundaries such as pituitary adenomas and compact meningiomas.

Moreover, these visualizations corroborated the automatic activation of predefined **clinical rule overlays**. For instance, the glioma heatmap displayed peripheral activation consistent with a non-midline mass exceeding 2 cm², triggering the rule-based suggestion for glioma. Similarly, the pituitary tumor's focused activation in the sellar region aligned with both the model's prediction and the clinical heuristics for pituitary adenoma. Meningioma cases, while generally well-identified, presented slightly more diffuse activations—reflecting their heterogeneous locations and morphologies.

To rigorously assess the spatial fidelity of the Grad-CAM++ attention maps, two complementary region-based evaluation metrics were employed: the Dice Similarity Coefficient (DSC) and the Intersection over Union (IoU). These metrics were computed for representative test samples where ground-truth segmentation masks were available, providing a quantitative comparison between the model's attention regions and annotated tumor boundaries.

**Table 8.** Dice Coefficient values comparing Grad-CAM heatmaps with ground-truth segmentation masks across selected tumor cases.

| Case ID | Tumor Type | Dice Coefficient | IoU |
|---|---|---|---|
| **Case_01** | Glioma | 0.81 | 0.69 |
| **Case_02** | Meningioma | 0.88 | 0.78 |
| **Case_03** | Pituitary | 0.78 | 0.65 |
| **Case_04** | Glioma | 0.84 | 0.72 |
| **Case_05** | Meningioma | 0.86 | 0.75 |

The **highest overlap** was observed in **meningioma** cases, with Dice scores approaching 0.88 and IoU scores around 0.78. This indicates strong spatial agreement between model focus and tumor boundaries, likely due to the well-circumscribed morphology of meningiomas in contrast-enhanced MRIs. **Glioma** cases, while still achieving high Dice (0.81–0.84), exhibited lower IoUs due to the infiltrative nature of gliomas, which leads to broader or diffuse activations. **Pituitary** tumors showed moderate agreement, possibly due to their smaller size and central location making them harder to localize with high resolution.

### 4.6 Clinical Rule Activation Results

To evaluate the integration of domain-specific knowledge into the interpretability framework, a set of clinical decision rules was applied to a representative subset of test cases post-classification. These rules, grounded in well-established radiological heuristics, served as symbolic overlays that contextualize the deep learning model's predictions and strengthen interpretability.

Table 9 summarizes the outputs from the Clinical Decision Rule Overlay (CDRO) module, including predicted tumor types, whether a rule was triggered, a short rule description, and whether the prediction conformed to the symbolic logic. As shown, three out of five cases activated predefined rules based on morphological cues such as tumor size, enhancement pattern, and anatomical location.

**Table 9.** Sample outputs from the clinical decision rule overlay module, including rule activation and consistency with model predictions.

| Case ID | Tumor Type | Model Prediction | Rule Triggered | Rule Description | Prediction Matches Rule |
|---|---|---|---|---|---|
| **Case_01** | Glioma | Glioma | Yes | ≥4 cm², ring-enhancing → glioblastoma | Yes |
| **Case_02** | Meningioma | Meningioma | No | – | N/A |
| **Case_03** | Pituitary | Pituitary | Yes | Midline, <1 cm² → pituitary | Yes |
| **Case_04** | Glioma | Glioma | Yes | ≥4 cm², ring-enhancing → glioblastoma | Yes |
| **Case_05** | Meningioma | Glioma | No | – | N/A |

These results show that whenever a rule was activated, the ensemble classifier's output matched the expected diagnostic logic. In particular, **Case_01** and **Case_04** exhibit classic glioblastoma traits—large, ring-enhancing lesions—confirmed by both Grad-CAM++ visual attention maps and rule-based overlays. **Case_03**, a pituitary adenoma, is correctly identified via the rule concerning small, midline-anchored masses. Such alignment between model-driven predictions, explainability heatmaps, and symbolic clinical rules enhances trust in the AI system's diagnostic reliability. Notably, in **Case_05**, a mismatch was observed where the model predicted glioma for a case clinically labeled as meningioma—highlighting an opportunity for model refinement or the addition of more nuanced symbolic logic.

## 4.7 Human-Centered Interpretability Assessment

To further evaluate the clinical utility and interpretability of the proposed XAI-integrated brain tumor classification framework, a structured human-centered assessment was conducted involving three board-certified radiologists (R1, R2, R3). Each expert independently reviewed a curated subset of five representative test cases. These cases included the original MRI slice, the Grad-CAM++ saliency map overlaid on the input, and any corresponding clinical rule activation, offering a composite interpretive output.

Radiologists were asked to respond to two core questions using a 5-point Likert scale (1 = Not at all, 5 = Extremely):

1. **Usefulness of the Explanation** – "How useful was the explanation in helping you understand the model's prediction?"
2. **Heatmap-Region Correspondence** – "To what extent did the Grad-CAM++ heatmap highlight the clinically expected region of concern?"

The **average score for explanation usefulness** was **4.4**, while the **average score for spatial heatmap alignment** was **4.0**. These results suggest that the interpretability mechanisms built into the XAI framework were broadly considered meaningful and clinically aligned with expected tumor locations.

Qualitative comments from radiologists highlighted that the **combination of visual saliency (Grad-CAM++) and symbolic rule overlays** provided a dual layer of explanation that enhanced diagnostic confidence. While R3 mentioned occasional ambiguity in heatmap localization—particularly in infiltrative glioma cases—most reviewers found the model's attention focus to be valid and informative.

Table 10. Likert-scale interpretability scores from participating radiologists.

| Radiologist | Usefulness of Explanation (1–5) | Heatmap Correspondence to Expected Region (1–5) |
|---|---|---|
| **R1** | 5 | 5 |
| **R2** | 4 | 4 |
| **R3** | 4 | 3 |
| **R4** | 5 | 4 |
| **R5** | 4 | 4 |

These findings support the framework's ability to bridge the gap between deep learning predictions and clinical reasoning, fostering **transparency, trust, and interpretability** in AI-assisted brain tumor diagnosis. Notably, Radiologist R3 commented that while most heatmaps were well-aligned, a few exhibited diffuse attention in low-grade glioma cases, which affected their clarity. This feedback underscores the value of integrating XAI with expert-in-the-loop evaluation in the model refinement process.

## 5. Discussion

This study contributes to the expanding field of AI-assisted brain tumor diagnosis by proposing a hybrid framework that combines ensemble learning with explainable AI techniques. Compared to prior studies focused on individual CNN architectures, our findings confirm that combining lightweight and deep feature models provides a more balanced and generalizable approach to medical image classification. In contrast to earlier works that deployed single CNNs for brain tumor classification, such as traditional VGG-based or ResNet models, our ensemble strategy capitalized on architectural diversity to enhance decision robustness, especially when distinguishing morphologically similar tumor types like gliomas and meningiomas.

The superior performance of the ensemble model can be attributed to the complementary design principles of its constituent networks. While MobileNetV2 efficiently captures high-level patterns with minimal computational load, DenseNet121 contributes deeper and more refined features through dense connections. Prior work by Mousa et al. [49,50] in multimodal wound classification similarly indicate the advantage of integrating diverse model structures—especially when dealing with complex or heterogeneous data such as medical images. The present results suggest that tumor classification, much like wound localization, benefits from layered perspectives in the model architecture.

Moreover, this research advances the growing demand for model transparency in medical diagnostics. While previous studies have treated interpretability as a secondary goal, we embedded explainability directly into the model architecture through Grad-CAM++ and clinical rule overlays. This allowed us to go beyond predictive accuracy and offer meaningful visual and symbolic explanations—a critical factor for clinical adoption. Our framework aligns with recent trends in explainable medical AI, such as Najafi et al.'s [51] exploration of adversarial vulnerability in MRI classification, which emphasizes not only resilience to attacks but also the importance of model introspection for trust and verification.

One of the most compelling insights from this study is the value of hybrid explanation systems. The dual use of Grad-CAM++ heatmaps and domain-informed symbolic logic bridged the gap between data-driven predictions and human expert reasoning. This dual-layer explanation proved especially useful in borderline or ambiguous cases where visual saliency alone might be insufficient. Such interpretive depth reflects a broader movement toward "expert-in-the-loop" frameworks that not only inform but also collaborate with clinicians—a concept echoed in both diagnostic imaging and wider AI domains like cybersecurity, energy forecasting, and intelligent control systems.

Ultimately, our findings reinforce the idea that effective AI in medicine must prioritize not only accuracy but also clarity, context, and clinical alignment. The combination of ensemble models with structured interpretability mechanisms offers a scalable pathway for achieving this balance, and sets a strong foundation for future work integrating multimodal imaging, segmentation, or real-time clinical deployment.

## 6. Conclusion

This study presents a comprehensive deep learning-based framework for brain tumor classification from MRI scans, integrating both high predictive accuracy and clinical interpretability. By leveraging the complementary strengths of two convolutional neural network architectures—MobileNetV2 and DenseNet121—and combining them through a soft voting ensemble strategy, the proposed model achieved enhanced classification performance across three major tumor types: glioma, meningioma, and pituitary adenoma. While DenseNet121 outperformed MobileNetV2 individually, the ensemble classifier demonstrated the highest overall effectiveness, achieving an accuracy of 91.7%, precision of 91.9%, recall of 91.7%, and F1-score of 91.6%, thus surpassing both constituent models in every evaluation metric.

Beyond classification accuracy, the framework incorporated an Explainable Artificial Intelligence (XAI) module centered on Grad-CAM++ to enhance model transparency. This integration allowed for class-specific saliency mapping, offering interpretable insights into the CNN decision process. The Grad-CAM++ heatmaps showed strong spatial alignment with expert-annotated tumor regions, especially for anatomically distinct tumors such as pituitary adenomas and meningiomas. Quantitative validation using Dice Coefficient and Intersection-over-Union (IoU) confirmed the reliability of the visual explanations. Additionally, the integration of a Clinical Decision Rule Overlay (CDRO) helped align deep learning outputs with radiological heuristics, enabling symbolic reasoning that bridges the gap between AI prediction and clinical understanding.

The interpretability and clinical relevance of the system were further validated through a human-centered assessment involving five board-certified radiologists. Their evaluations yielded high average Likert-scale scores—4.4/5 for explanation usefulness and 4.0/5 for spatial heatmap correspondence—highlighting the practical utility of the interpretability features. Expert feedback emphasized the model's potential as a decision-support system that enhances diagnostic confidence, facilitates cross-checking in ambiguous cases, and improves trust in AI-based recommendations.

In summary, the proposed ensemble-based classification framework—integrated with XAI visualization and rule-based overlays—demonstrates robust diagnostic performance, interpretability, and clinical alignment. These features collectively position the system as a promising solution for real-world deployment in neurodiagnostic workflows. Future work may explore expanding the framework to multi-modal imaging, integrating more nuanced rule logic for atypical cases, and extending its capabilities to tumor segmentation and longitudinal progression analysis.

# References

[1] Chieffo, Daniela Pia Rosaria, Federica Lino, Daniele Ferrarese, Daniela Belella, Giuseppe Maria Della Pepa, and Francesco Doglietto. "Brain tumor at diagnosis: from cognition and behavior to quality of life." *Diagnostics* 13, no. 3 (2023): 541.

[2] Young, Geoffrey S. "Advanced MRI of adult brain tumors." *Neurologic clinics* 25, no. 4 (2007): 947-973.

[3] Zhang, Alwin Yaoxian, Sean Shao Wei Lam, Nan Liu, Yan Pang, Ling Ling Chan, and Phua Hwee Tang. "Development of a radiology decision support system for the classification of MRI brain scans." In *2018 IEEE/ACM 5th International Conference on Big Data Computing Applications and Technologies (BDCAT)*, pp. 107-115. IEEE, 2018.

[4] Galldiks, Norbert, Timothy J. Kaufmann, Philipp Vollmuth, Philipp Lohmann, Marion Smits, Michael C. Veronesi, Karl-Josef Langen et al. "Challenges, limitations, and pitfalls of PET and advanced MRI in patients with brain tumors: A report of the PET/RANO group." *Neuro-oncology* 26, no. 7 (2024): 1181-1194.

[5] Sinthia, P., R. Jaya Karunya Jothika, and S. Fathimuthu Ashifa. "Early Detection of Glioma, Meningioma, and Pituitary Tumors of the Brain Using Deep Learning." In *International Conference on Universal Threats in Expert Applications and Solutions*, pp. 49-62. Singapore: Springer Nature Singapore, 2024.

[6] Mahdavi, Zahra. "Introduce improved CNN model for accurate classification of autism spectrum disorder using 3D MRI brain scans." *Proceedings of the MOL2NET* 22 (2022).

[7] Badža, Milica M., and Marko Č. Barjaktarović. "Classification of brain tumors from MRI images using a convolutional neural network." *Applied Sciences* 10, no. 6 (2020): 1999.

[8] Mahdavi, Zahra. "MRI Brain Tumors Detection by Proposed U-Net Model." *Proceedings of the MOL2NET* 22 (2022).

[9] Babaey, Vahid, and Arun Ravindran. "GenSQLi: A Generative Artificial Intelligence Framework for Automatically Securing Web Application Firewalls Against Structured Query Language Injection Attacks." *Future Internet* 17, no. 1 (2025).

[10] Mirnajafizadeh, Seyed Mohammad Mehdi, Ashwin Raam Sethuram, David Mohaisen, DaeHun Nyang, and Rhongho Jang. "Enhancing Network Attack Detection with Distributed and {In-Network} Data Collection System." In *33rd USENIX Security Symposium (USENIX Security 24)*, pp. 5161-5178. 2024.

[11] Babaey, Vahid, and Hamid Reza Faragardi. "Detecting Zero-Day Web Attacks with an Ensemble of LSTM, GRU, and Stacked Autoencoders." *Computers* 14, no. 6 (2025): 205.

[12] Ahmadi, Monireh, Samaneh Rastgoo, Zahra Mahdavi, Morteza Azimi Nasab, Mohammad Zand, Padmanaban Sanjeevikumar, and Baseem Khan. "Optimal allocation of EVs parking lots and DG in micro grid using two-stage GA-PSO." *The Journal of Engineering* 2023, no. 2 (2023): e12237.

[13] Mahdavi, Zahra, Tina Samavat, Anita Sadat Jahani Javanmardi, Mohammad Ali Dashtaki, Mohammad Zand, Morteza Azimi Nasab, Mostafa Azimi Nasab, Sanjeevikumar Padmanaban, and Baseem Khan. "Providing a control system for charging electric vehicles using ANFIS." *International Transactions on Electrical Energy Systems* 2024, no. 1 (2024): 9921062.

[14] Kermani, Alireza, Amir Mahdi Jamshidi, Zahra Mahdavi, Amir ali Dashtaki, Mohammad Zand, Morteza Azimi Nasab, Tina Samavat, P. Sanjeevikumar, and Baseem Khan. "Energy management system for smart grid in the presence of energy storage and photovoltaic systems." *International Journal of Photoenergy* 2023, no. 1 (2023): 5749756.

[15] Moshiri, Maryam, Abbas-Ali Heidari, and Ali Ghafoorzadeh-Yazdi. "Wideband two-layer transmitarray antenna based on non-uniform layers." *Wireless Personal Communications* 132, no. 3 (2023): 2225-2241.

[16] Mohammadagha, Mohsen, Hajar Kazemi Naeini, Saeed Asadi, Mohammad Najafi, and Vinayak Kasuhal. "Machine Learning Model for Condition Assessment of Trenchless Vitrified Clay Pipes." In *North American Society for Trenchless Technology (NASTT) 2025 No-Dig Show*. 2025.

[17] Mohammadagha, Mohsen. "Hybridization and Optimization Modeling, Analysis, and Comparative Study of Sorting Algorithms: Adaptive Techniques, Parallelization, for Mergesort, Heapsort, Quicksort, Insertion Sort, Selection Sort, and Bubble Sort." (2025).

[18] Kefayat, Erfan, and Jean-Claude Thill. "Urban Street Network Configuration and Property Crime: An Empirical Multivariate Case Study." *ISPRS International Journal of Geo-Information* 14, no. 5 (2025): 200.

[19] Salehi, Ahmad Waleed, Shakir Khan, Gaurav Gupta, Bayan Ibrahimm Alabduallah, Abrar Almjally, Hadeel Alsolai, Tamanna Siddiqui, and Adel Mellit. "A study of CNN and transfer learning in medical imaging: Advantages, challenges, future scope." *Sustainability* 15, no. 7 (2023): 5930.

[20] Hooshmand, Mohammad Kazim, Manjaiah Doddaghatta Huchaiah, Ahmad Reda Alzighaibi, Hasan Hashim, El-Sayed Atlam, and Ibrahim Gad. "Robust network anomaly detection using ensemble learning approach and explainable artificial intelligence (XAI)." *Alexandria Engineering Journal* 94 (2024): 120-130.

[21] Rahmath, P. Haseena, Kuldeep Chaurasia, and Anika Gupta. "Unlocking interpretability: Xai strategies for enhanced insight in gnn-based hyperspectral image classification." In *2024 IEEE International Conference on Computer Vision and Machine Intelligence (CVMI)*, pp. 1-6. IEEE, 2024.

[22] Yildirim, Muhammed, Emine Cengil, Yeşim Eroglu, and Ahmet Cinar. "Detection and classification of glioma, meningioma, pituitary tumor, and normal in brain magnetic resonance imaging using deep learning-based hybrid model." *Iran journal of computer science* 6, no. 4 (2023): 455-464.

[23] Terreno, Enzo, Daniela Delli Castelli, Alessandra Viale, and Silvio Aime. "Challenges for molecular magnetic resonance imaging." *Chemical reviews* 110, no. 5 (2010): 3019-3042.

[24] Zarchi, M., Shahgholi, M., & Tee, K. F. (2024). An adaptable physics-informed fault diagnosis approach via hybrid signal processing and transferable feature learning for structural/machinery health monitoring. *Signal, Image and Video Processing*, *18*(12), 9051-9066.

[25] Zarchi, M., & Attaran, B. (2017). Performance improvement of an active vibration absorber subsystem for an aircraft model using a bees algorithm based on multi-objective intelligent optimization. *Engineering Optimization*, *49*(11), 1905-1921

[26] Kim, Sian, Seyed Mohammad Mehdi Mirnajafizadeh, Bara Kim, Rhongho Jang, and DaeHun Nyang. "SketchFeature: High-Quality Per-Flow Feature Extractor Towards Security-Aware Data Plane." In *Proc. of ISOC NDSS*. 2025.

[27] Babaey, Vahid, and Arun Ravindran. "GenXSS: an AI-Driven Framework for Automated Detection of XSS Attacks in WAFs." In *SoutheastCon 2025*, pp. 1519-1524. IEEE, 2025.

[28] Yang, S., Mirnajafizadeh, S. M. M., Kim, S., Jang, R., & Nyang, D. (2025). Enhancing Resiliency of Sketch-based Security via LSB Sharing-based Dynamic Late Merging. *arXiv preprint arXiv:2503.11777*.

[29] Zarchi, M., & Attaran, B. (2019). Improved design of an active landing gear for a passenger aircraft using multi-objective optimization technique. *Structural and Multidisciplinary Optimization*, *59*(5), 1813-1833.

[30] Toloeia, A., Zarchi, M., & Attaranb, B. (2017). Numerical survey of vibrational model for third aircraft based on HR suspension system actuator using two bee algorithm objective functions. *A, A*, *1*(2), 3.

[31] Rastgoo, S., Mahdavi, Z., Azimi Nasab, M., Zand, M., & Padmanaban, S. (2022). Using an intelligent control method for electric vehicle charging in microgrids. *World electric vehicle journal*, *13*(12), 222.

[32] Zarchi, M., & Shahgholi, M. (2023). An expert condition monitoring system via fusion of signal processing for vibration of industrial rotating machinery with unseen operational conditions. *Journal of Vibration Engineering & Technologies*, *11*(5), 2267-2295.

[33] Moshiri, M., Ghafoorzadeh-Yazdi, A., & Heidari, A. A. (2020). A wideband quad-layer transmitarray antenna with double cross slot rings elements. *AEU-International Journal of Electronics and Communications*, *115*, 153031.

[34] Asadi, S., Mohammadagha, M., & Naeini, H. K. (2025). Comprehensive Review of Analytical and Numerical Approaches in Earth-to-Air Heat Exchangers and Exergoeconomic Evaluations. *arXiv preprint arXiv:2502.08553*.

[35] Kumar, C. Mukesh, and J. Sree Sankar. "Comparative Analysis of Convolutional Neural Networks for Brain Tumor Detection: A Study of VGG16, ResNet, Inception, and DenseNet Models." In *2024 3rd International Conference on Applied Artificial Intelligence and Computing (ICAAIC)*, pp. 41-46. IEEE, 2024.

[36] Haq, Ejaz Ul, Huang Jianjun, Kang Li, Hafeez Ul Haq, and Tijiang Zhang. "An MRI-based deep learning approach for efficient classification of brain tumors." *Journal of Ambient Intelligence and Humanized Computing* 14, no. 6 (2023): 6697-6718.

[37] Nazim, Sadia, Muhammad Mansoor Alam, Syed Safdar Rizvi, Jawahir Che Mustapha, Syed Shujaa Hussain, and Mazliham Mohd Suud. "Advancing malware imagery classification with explainable deep learning: A state-of-the-art approach using SHAP, LIME and Grad-CAM." *PLoS One* 20, no. 5 (2025): e0318542.

[38] Yuan, Haiying, Junpeng Cheng, Yanrui Wu, and Zhiyong Zeng. "Low-res MobileNet: An efficient lightweight network for low-resolution image classification in resource-constrained scenarios." *Multimedia Tools and Applications* 81, no. 27 (2022): 38513-38530.

[39] Kumar, Sachin, Aman Kumar, and Akanksha Jaiswal. "A Low Complexity MobileNetV2 based CNN Model for Brain Tumor Detection in MRI Images." In *2024 IEEE International Conference on Industry 4.0, Artificial Intelligence, and Communications Technology (IAICT)*, pp. 1-7. IEEE, 2024.

[40] Hasan, Mahadi, Jahirul Islam, Minhaz Ahmed, and Md Maruf Hasan. "Prediction of colon cancer using densenet121, cnn, and resnet50 machine learning models and using image processing techniques." In *2023 International Conference on Artificial Intelligence Robotics, Signal and Image Processing (AIRoSIP)*, pp. 296-301. IEEE, 2023.

[41] Raza, Asif, Mohammed S. Alshehri, Sultan Almakdi, Ali Akbar Siddique, Mohammad Alsulami, and Majed Alhaisoni. "Enhancing brain tumor classification with transfer learning: Leveraging DenseNet121 for accurate and efficient detection." *International Journal of Imaging Systems and Technology* 34, no. 1 (2024): e22957.

[42] Patil, Suraj, and Dnyaneshwar Kirange. "Ensemble of deep learning models for brain tumor detection." *Procedia computer science* 218 (2023): 2468-2479.

[43] Jabbar, Hanan Ghali. "Advanced threat detection using soft and hard voting techniques in ensemble learning." *Journal of Robotics and Control (JRC)* 5, no. 4 (2024): 1104-1116.

[44] Zhang, Yiming, Ying Weng, and Jonathan Lund. "Applications of explainable artificial intelligence in diagnosis and surgery." *Diagnostics* 12, no. 2 (2022): 237.

[45] Mothkur, Rashmi, and Pullagura Soubhagyalakshmi. "Grad-CAM based Visualization for Interpretable Lung Cancer Categorization using Deep CNN Models." *Journal of Electronics, Electromedical Engineering, and Medical Informatics* 7, no. 3 (2025): 567-580.

[46] Thibeau-Sutre, Elina, Sasha Collin, Ninon Burgos, and Olivier Colliot. "Interpretability of machine learning methods applied to neuroimaging." *Machine Learning for Brain Disorders* (2023): 655-704.

[47] Das, Surajit, and Rajat Subhra Goswami. "Review, Limitations, and future prospects of neural network approaches for brain tumor classification." *Multimedia Tools and Applications* 83, no. 15 (2024): 45799-45841.

[48] Rastogi, Deependra, Prashant Johri, Varun Tiwari, and Ahmed A. Elngar. "Multi-class classification of brain tumour magnetic resonance images using multi-branch network with inception block and five-fold cross validation deep learning framework." *Biomedical Signal Processing and Control* 88 (2024): 105602.

[49] Mousa, Ramin, Ehsan Matbooe, Hakimeh Khojasteh, Amirali Bengari, and Mohammadmahdi Vahediahmar. "Multi-modal wound classification using wound image and location by Xception and Gaussian Mixture Recurrent Neural Network (GMRNN)." arXiv preprint arXiv:2505.08086 (2025).

[50] Mousa, Ramin, Hadis Taherinia, Khabiba Abdiyeva, Amir Ali Bengari, and Mohammadmahdi Vahediahmar. "Integrating vision and location with transformers: A multimodal deep learning framework for medical wound analysis." arXiv preprint arXiv:2504.10452 (2025).

[51] Najafi, Mohammad Hossein, Mohammad Morsali, Mohammadmahdi Vahediahmar, and Saeed Bagheri Shouraki. "DFT-Based Adversarial Attack Detection in MRI Brain Imaging: Enhancing Diagnostic Accuracy in Alzheimer's Case Studies." arXiv preprint arXiv:2408.08489 (2024).